# Powering the Future of IoT: Federated Learning for Optimized Power Consumption and Enhanced Privacy

Ghazaleh Shirvani
ghazalehshirvani@cmail.carleton.ca
Carleton University
Ottawa, Ontario, Canada

Saeid Ghasemshirazi
saeidghasemshirazi@cmail.carleton.ca
Carleton University
Ottawa, Ontario, Canada

## ABSTRACT

The widespread use of Internet of Things has led to the development of large amounts of perception data, making it necessary to develop effective and scalable data analysis tools. Federated Learning emerges as a promising paradigm to address the inherent challenges of power consumption and data privacy in IoT environments. This paper explores the transformative potential of FL in enhancing the longevity of IoT devices by mitigating power consumption and enhancing privacy and security measures. We delve into the intricacies of FL, elucidating its components and applications within IoT ecosystems. Additionally, we discuss the critical characteristics and challenges of IoT, highlighting the need for such machine learning solutions in processing perception data. While FL introduces many benefits for IoT sustainability, it also has limitations. Through a comprehensive discussion and analysis, this paper elucidates the opportunities and constraints of FL in shaping the future of sustainable and secure IoT systems. Our findings highlight the importance of developing new approaches and conducting additional research to maximise the benefits of FL in creating a secure and privacy-focused IoT environment.

## 1 INTRODUCTION

The Internet of Things (IoT) has witnessed exponential growth in recent years, with reports indicating that the average household now hosts at least seven IoT devices [51]. These devices span a diverse range, from wearable gadgets and smartphones to laptops, collectively forming an intricate network of interconnected entities. The fundamental principle of IoT devices lies in their ability to connect and collaborate, leveraging shared data to gain insights and drive intelligent decision-making processes [3].

The primary goal of IoT devices goes beyond simple connectivity; it includes the process of obtaining, analyzing, and interpreting data to comprehend larger concepts and patterns [37]. Thus, this is how IoT devices play a pivotal role in shaping our interconnected world, either by analyzing consumer behavior to provide personalized support or by identifying important health indicators to encourage appropriate actions. Therefore, in order to understand and make decisions based on these patterns, advance automated approaches (i.e. machine learning methodologies, data mining) are necessary when dealing with such large amounts of data [45].

Despite the great potential of IoT in various domains, it faces inherent challenges related to resource constraints, particularly in computation and power consumption. These constraints pose significant obstacles in deploying machine learning algorithms effectively, which typically demand substantial computational and power resources. Additionally, as IoT devices increasingly find applications in sensitive and critical contexts, ensuring the privacy and security of data becomes vital.

In response to these challenges, Federated Learning (FedML) has emerged as a groundbreaking paradigm in distributed machine learning. Research has been employed in FL to tackle the unique requirements and constraints of IoT ecosystems. In 2016 [49] Introduced FedML that allows for collaborative model training on decentralized devices. This enables the creation of a globally shared model without revealing raw data or sacrificing user privacy. By transmitting only model parameters rather than raw data, FedML significantly reduces communication overhead and, consequently, power consumption, making it particularly suitable for energy-constrained IoT environments [29].

This study seeks to underscore the pivotal role of Federated Learning in enhancing privacy, reducing power consumption, and fostering a sustainable IoT ecosystem. Through a comprehensive analysis of existing literature, this study aim to elucidate the potential of FedML in addressing the evolving challenges of IoT deployments and charting new avenues for future research.

**Research Questions.** In order to provide guidance for the advancement of more sustainable IoT ecosystems, the present study aims to investigate the subsequent research questions:

**(1)** How can Federated Learning (FedML) contribute to reducing power consumption in IoT devices? This question aims to explore the potential benefits of FedML in optimizing power usage, thus extending longevity of IoT devices.

**(2)** What are the existing algorithms and methodologies for optimizing Federated Learning in energy-constrained IoT environments? This question focuses on reviewing and analyzing the current state-of-the-art algorithms and techniques that aim to enhance the efficiency and effectiveness of FedML in IoT settings.

**(3)** How does Federated Learning address the privacy and security concerns associated with IoT devices? This question seeks to investigate the mechanisms and approaches employed by FedML to safeguard user privacy and enhance security, thereby reducing the likelihood of premature device decommissioning due to perceived malfunctions or security breaches.

**(4)** How can Federated Learning contribute to creating a more sustainable and resilient IoT ecosystem?

## 2 BACKGROUND

In this part, we provide a concise overview of the relevant background for the main concepts.

### 2.1 IoT Ecosystem

The Internet of Things ecosystem encompasses a diverse array of devices designed to connect, communicate, and collaborate to deliver intelligent functionalities across various domains. These devices can be broadly categorized based on their applications, form factors, and power requirements. Wearable devices represent a prominent category within the IoT landscape, comprising devices like smartwatches, fitness trackers, and health monitors. These devices are engineered to be lightweight, portable, and energy-efficient to ensure extended battery life, thereby enhancing user convenience and usability [57].

Smart home devices constitute another significant segment of IoT devices, including smart thermostats, smart lights, and smart locks. Although these devices are usually connected to the main power supply, it is crucial to optimise their power consumption in order to reduce energy usage and operating expenses without sacrificing functionality [41]. Healthcare IoT devices, such as medical monitors, remote patient monitoring systems, and drug delivery systems, are designed to facilitate remote healthcare monitoring and management. Operating on battery power, these devices necessitate low power consumption to ensure reliable and continuous operation, especially in critical healthcare scenarios [43].

Smart transportation is another domain that is witnessing the proliferation of IoT devices, including connected vehicles, traffic monitoring systems, and smart infrastructure. The power requirements of these devices vary significantly based on their applications, ranging from battery-powered sensors to electric vehicle charging stations, each have unique power consumption profiles. Virtual Power Plants (VPPs), which are a cutting-edge solution for controlling distributed energy resources like solar panels, batteries, and wind turbines, are encompassed within the scope of the Internet of Things [9].

### 2.2 IoT Perception Data Analysis

The proliferation of Internet of Things has resulted in unprecedented growth in the volume, variety, and velocity of data. This vast amount of data, known as perception data, has the ability to provide vital insights and facilitate intelligent decision-making in several fields [38]. IoT perception data involves several sorts of data, such as sensor data, image and video data, audio data, and location data.

Data is collected from various embedded sensors, which offer valuable information about the physical conditions and activities in the device's surroundings. Visual analytics, object recognition, and surveillance applications are made possible by the use of integrated cameras and imaging devices to record image and video data. Audio data, obtained from microphones and audio sensors, enables the application of speech recognition, sound classification, and ambient noise monitoring. In addition, location data obtained from GPS-enabled devices provides details on the device's specific geographic location and patterns of movement [20].

In terms of applications, IoT perception data analysis enables a wide range of creative solutions in many areas. For instance, wearable sensor data analysis in smart healthcare allows for ongoing health monitoring, identification of abnormalities, and prediction of potential risks, thereby facilitating personalised and proactive healthcare services [58]. In industrial IoT applications, sensor data analytics support predictive maintenance, anomaly detection, and process optimization, enhancing operational efficiency and minimizing downtime. Similarly, in smart agriculture, environmental and soil sensor data analysis facilitates to optimise irrigation, monitor crop health, and anticipate weather conditions. This ultimately leads to increased agricultural output and sustainability. Lastly, in smart cities, the analysis of traffic, environmental, and energy consumption data enables optimized urban planning, enhanced public safety, and sustainable development within smart city initiatives [2].

Despite the vast potential of IoT perception data, several challenges hinder its effective analysis and utilization. One of the primary challenges is the heterogeneity and complexity of IoT data, characterized by varying formats, structures, and characteristics. This diversity complicates data integration and analysis, necessitating sophisticated data processing techniques capable of handling heterogeneous data streams. Moreover, the volume of data generated by IoT devices can overwhelm traditional data processing and analytics systems, requiring scalable and efficient data analysis approaches [1]. Additionally, many IoT applications demand realtime or near-real-time data analysis to facilitate prompt decision-making and reaction, which poses challenges with achieving high latency requirements. Furthermore, the sensitive and personal nature of IoT perception data raises concerns about data privacy, confidentiality, and security, emphasizing the need for robust data protection mechanisms to safeguard user privacy and comply with regulatory requirements [62].

Having a deep understanding of IoT perception data analysis is essential for unlocking the full potential of IoT devices and fostering innovation in different fields. In following, by addressing the challenges associated with data variety, volume, real-time processing, and privacy, we can discover new avenues for revolutionary advancements in IoT-enabled sustainability and longevity solutions.

## 2.3 IoT Characteristics and Challenges

As discussed, IoT devices constitute a diverse ecosystem of interconnected devices ranging from sensors and actuators to wearables and smart appliances. These devices, often characterized by their small form factor and connectivity to the internet. Nevertheless, the unique characteristics of IoT devices pose significant challenges for ensuring security and privacy in this rapidly growing landscape [5].

Here are some characteristics of IoT:

**(1) Resource Limitation:** IoT devices often operate with constrained computational resources, including limited processing power, memory, and energy. Due to the large amount of perceptual data produced by IoT devices and the urgent requirement for data analysis, machine learning algorithms are essential. Consequently, IoT resources constraints necessitate the development of lightweight and energy-efficient machine learning algorithms tailored for edge computing environments [28].

**(2) Heterogeneity:** The diverse range of IoT devices, sensors, protocols, and data formats contribute to the heterogeneity of IoT ecosystems, posing challenges for data integration, interoperability, and standardization [54]. In other words, devices collect data that is not identically distributed (IID) across the network. This means the data on each device may not be representative of the overall data distribution. Machine learning techniques capable of handling heterogeneous data streams and adapting to diverse IoT environments are essential for ensuring connectivity, collaboration, and data exchange across disparate devices and platforms.

**(3) Privacy Concerns:** IoT deployments in critical applications, such as Virtual Power Plants (VPPs), hospitals, and farms, have tight privacy requirements to protect sensitive information and comply with regulatory guidelines. However, the decentralized and distributed nature of IoT devices necessitates the use of distributed algorithms, which pose significant challenges [33]. Thus, employing machine learning algorithms that contributes to user privacy are essential for this characteristic of IoT devices.

## 2.4 Power Requirements of IoT Devices

Having a deep understanding of the power requirements of IoT devices is crucial in order to develop sustainable IoT environment and maximize their performance. Several factors influence the power consumption of IoT devices, including battery life, processing capabilities, communication protocols, sensing and data collection mechanisms, and the operational environment.

Battery life is a critical consideration for many IoT devices, particularly wearable and healthcare devices operating on battery power. Extending battery life is essential to ensure uninterrupted operation, enhance user experience, minimize the frequency of battery replacements, and thus having a more a long-lasting IoT device [10]. The table 1 provides a general overview of the power requirements for various categories of IoT devices. It's important to note that mentioned power consumption ranges are estimates and the actual power consumption of a device can vary depending on factors like sensor usage, communication frequency, and processing power.

The processing capabilities of IoT devices can significantly impact power consumption. Devices equipped with advanced processors and functionalities often consume more power. Hence, optimizing algorithms and hardware configurations are essential to minimize power consumption without compromising performance. Communication protocols utilized by IoT devices, such as Wi-Fi, Bluetooth, Zigbee, and LoRaWAN, also play a crucial role in determining power consumption [23]. The choice of communication protocol can influence power efficiency, with some protocols optimized for low power consumption and others designed for higher data rates [22].

Furthermore, the sensing and data collection mechanisms employed by IoT devices, including temperature sensors, motion sensors, and environmental sensors, can contribute to power consumption, particularly when collecting and transmitting data at frequent intervals. Lastly, the operational environment, encompassing factors like temperature, humidity, and physical conditions, can affect the power requirements of IoT devices. Devices deployed in harsh or challenging environments may require additional power for protection, maintenance, and optimal performance [22].

## 2.5 Traditional Machine Learning for IoT

In the realm of Internet of Things security, traditional machine learning solutions have been extensively explored to address various security and privacy concerns. These solutions typically involve a centralized approach, wherein data collected from IoT devices is transmitted to a central server for aggregation and model training. However, while traditional ML approaches offer certain advantages, they also present notable limitations when applied to IoT security and privacy [55].

Traditional ML solutions for IoT security and privacy often rely on centralized data aggregation and model training. This approach entails transmitting data from IoT devices to a central server, where it is aggregated and utilized to train machine learning models. These models are subsequently deployed to detect anomalies, classify events/patterns (e.g. behavioral patterns), or identify security threats within the IoT ecosystem [63].

One of the notable advantages of ML solutions lies in their predictive capabilities. By analyzing historical data collected from IoT devices, ML models can effectively predict potential security threats or anomalies within IoT environments. Moreover, the centralized management of model training facilitates easier maintenance and updates of machine learning models, enhancing scalability and adaptability to evolving security threats [32].

However, despite these advantages, traditional ML solutions for IoT security and privacy are not without their drawbacks. Primary concern among these is the inherent privacy concern associated with centralized data aggregation. The transmission of sensitive information from IoT devices to a central server increases the risk of data breaches or unauthorized access, compromising the privacy of user data [32].

Furthermore, the centralized nature of traditional ML solutions results in significant communication overhead. Transmitting large volumes of data from IoT devices to a central server poses challenge such as increased bandwidth consumption and latency in resource-constrained IoT environments. Additionally, as the number of IoT devices continues to grow, centralized ML solutions struggle to scale effectively, leading to bottlenecks and performance degradation.

**Table 1: Typical Power Consumption Ranges of IoT Devices**

| Category | Typical Power Consumption Range | Example Devices |
|---|---|---|
| Low-power sensors | Microwatts (µW) to Milliwatts (mW) | Temperature sensors, humidity sensors, motion detectors |
| Wearable devices | Milliwatts (mW) to Hundreds of Milliwatts (mW) | Fitness trackers, smartwatches |
| Smart home devices | Tens of Milliwatts (mW) to Watts (W) | Smart thermostats, smart bulbs, smart plugs |
| Industrial IoT devices | Hundreds of Milliwatts (mW) to Tens of Watts (W) | Asset trackers, industrial sensors, actuators |
| Smartphones and tablets | Hundreds of Milliwatts (mW) to Watts (W) | While not strictly IoT devices, smartphones and tablets are often used to interact with IoT ecosystems and can have a significant impact on overall power consumption |

Another limitation of traditional ML solutions is their dependency on continuous connectivity between IoT devices and the central server. Disruptions in network connectivity or server downtime can adversely impact the effectiveness of ML-based security measures, leaving IoT ecosystems vulnerable to security threats [2]. All in all, considering the constraints of traditional machine learning in the IoT environment, there is an urgent requirement for new methods to address these limitations while ensuring data privacy and optimising resource usage.

## 2.6 Motivation for Federated Learning in IoT

In light of the challenges posed by traditional machine learning approaches in securing IoT devices, Federated Learning emerges as a compelling alternative that addresses several key drawbacks inherent in centralized ML solutions.

**Resource Limitation.** One of the primary drawbacks of traditional ML solutions for IoT security is the resource limitation imposed by IoT devices, including constraints on power and computational capacity. However, in the FL approach, this limitation is mitigated as each device is only responsible for processing a small set of data and participating in model training. By distributing the computational workload across multiple devices and leveraging local computation, FL minimizes the burden on individual devices, ensuring efficient resource utilization and preserving device performance [24].

**Power Consumption.** Traditional ML solutions often entail centralized data aggregation and model training, resulting in significant communication overhead and increased power consumption, which can adversely impact the battery life of IoT devices. In contrast, FL offers a more energy-efficient approach by reducing both the volume of data transmitted and the frequency of communications between devices and the central server. With smaller computations and fewer communications, FL prolongs the lifespan of batteries in IoT devices and minimizes overall power consumption, thereby enhancing the sustainability of IoT ecosystems [19].

**Flexibility in Connectivity.** Another advantage of FL in the context of IoT environments is its flexibility in accommodating variations in device connectivity. Unlike traditional ML solutions, which require continuous connectivity between IoT devices and the central server for model training, FL allows devices to participate in training asynchronously, without the need for synchronized communication. This asynchronous training mechanism enables devices to contribute to the model at their convenience, optimizing power consumption by minimizing idle time and ensuring efficient utilization of network resources.

## 2.7 Key Components of Federated Learning

Federated Learning operates on a unique architecture (Figure 1) that distinguishes it from traditional centralized machine learning systems. To understand FL's functionality better, it's crucial to delve into its core components: the Local Model, the Global Model, and the Aggregation Function.

**Local Model.** The Local Model refers to the machine learning model that resides on each participating device or edge node within the federated network. Unlike centralized systems where data is sent to a central server for processing, in FL, the Local Model is trained locally on each device using its respective data. This decentralized approach ensures that raw data remains on the device, thereby preserving data privacy and reducing the risks associated with data transmission [60].

Each Local Model learns from its locally stored data and generates updates or gradients, which are subsequently shared with the central server or aggregator during the aggregation phase. The Local Model serves as the foundation for personalized learning on individual devices, capturing device-specific insights and patterns without exposing sensitive data.

**Global Model.** The Global Model represents the comprehensive machine learning model that aggregates information from all Local Models across the federated network. The primary objective of the Global Model is to synthesize the collective knowledge from individual devices to generate a comprehensive and refined model that encapsulates the broader patterns and trends present in the data.

During the training process, the Global Model iteratively integrates the updates received from the Local Models, enabling collaborative learning while maintaining data privacy. This collaborative approach allows the Global Model to benefit from the diverse and decentralized data sources, resulting in a more robust and generalized model that reflects the collective intelligence of the network [21].

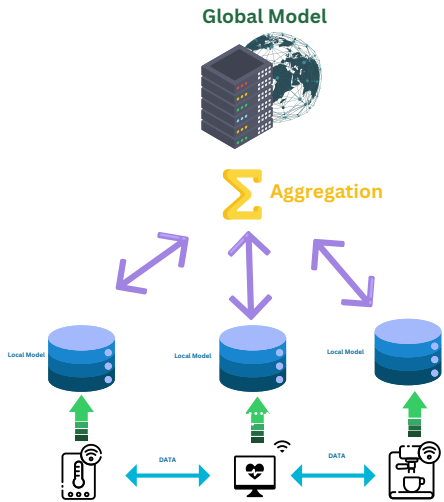

**Figure 1: Key Components of Federated Learning**

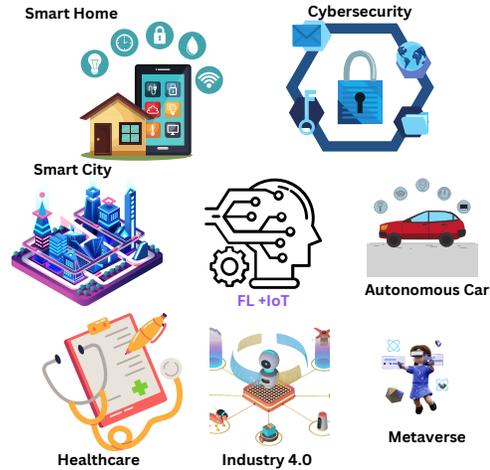

**Figure 2: Applications of Federated Learning for IoT Devices.**

**Aggregation Function.** The Aggregation Function plays a pivotal role in federated learning by facilitating the combination of updates or gradients from multiple Local Models to update the Global Model. Various aggregation functions can be employed based on the specific requirements and characteristics of the federated learning task.

Common aggregation functions include averaging, weighted averaging, and median-based aggregation, each offering unique advantages and trade-offs in terms of convergence speed, accuracy, and robustness. The selection of an aggregation function significantly impacts the efficiency and efficacy of the federated learning process. Therefore, it is crucial to choose an appropriate function that is specifically adapted to the requirements of the application [26].

*Take-away.* The Local Model, Global Model, and Aggregation Function collectively form the foundational elements of federated learning, enabling collaborative and privacy-preserving machine learning across distributed IoT devices. Understanding these components is crucial for grasping the mechanics of federated learning and its potential applications in various IoT scenarios.

## 3 INTRODUCTION TO FEDERATED IOT

Federated Learning has emerged as a promising approach to address privacy concerns and scalability issues associated with traditional machine learning systems. FL provides a decentralised training procedure where numerous edge devices collaborate to learn a global model. This collaboration occurs while the devices maintain their raw data locally safe and protected. This paradigm shift enables data privacy preservation, as sensitive information remains on the device; Hence, reduces the risks associated with centralized data aggregation.

In the context of the IoT, where an increasingly vast array of interconnected devices generates immense volumes of data, federated learning presents a compelling solution. IoT devices, ranging from sensors and actuators to wearables and smart appliances, collect data from their surroundings and perform various tasks autonomously or in conjunction with other devices. However, transmitting raw sensor data to centralized servers for training conventional machine learning models raises privacy, security, and bandwidth concerns [36].

Federated learning addresses these challenges by allowing IoT devices to collaboratively learn models without compromising data privacy. By leveraging the computational capabilities of edge devices, federated learning enables localized model training while aggregating global knowledge across the network. This distributed approach minimizes communication overhead and reduces reliance on centralized infrastructure, making federated learning well-suited for resource-constrained IoT environments [39].

The applications of federated learning in IoT are diverse and impactful. From enhancing predictive maintenance and anomaly detection in industrial IoT systems to facilitating personalized healthcare monitoring and smart home automation, federated learning empowers IoT devices to learn from each other's experiences while preserving data privacy. Moreover, federated learning enables continual model improvement through iterative updates, ensuring adaptability to dynamic IoT environments and evolving user preferences [4].

As federated learning continues to evolve and gain traction, it promises to revolutionize how IoT systems leverage machine learning while safeguarding data privacy and optimizing resource utilization. However, challenges such as model synchronization, heterogeneity of devices, and security vulnerabilities necessitate careful consideration and further research to fully realize the potential of federated learning in IoT applications [35].

In this paper, we delve deeper into the implications of federated learning on IoT security and power consumption, exploring its benefits, challenges, and future directions. By examining the intersection of federated learning, IoT security, and power consumption, we aim to provide insights into the opportunities and constraints of this transformative approach in shaping the future of IoT systems.

# 4 TRIANGLE OF FL, SECURITY, AND POWER CONSUMPTION

The intersection of Federated Learning, Internet of Things security, and power consumption represents a critical focus that has gained significant attention in recent years. [44]. As IoT devices proliferate across various domains, ensuring robust security measures and optimizing power consumption have become paramount concerns. Federated Learning emerges as a transformative approach that promises to address these challenges by offering privacy-preserving, decentralized machine learning capabilities tailored for resource-constrained IoT environments.

**Privacy-Preserving FL.** Privacy preservation stands as a fundamental requirement in IoT deployments, given the sensitive nature of the data generated and processed by IoT devices. Conventional centralised machine learning models have substantial privacy concerns since they often require gathering raw data in a single server, which might potentially expose sensitive information to unauthorised access or security breaches. In contrast, Federated Learning facilitates decentralised machine learning, allowing IoT devices to collaboratively learn a global model without sharing raw data. This paradigm shift ensures data privacy by design, as sensitive information remains localized on the device, thereby reducing the exposure to privacy breaches and unauthorized data access [17].

**Security Enhancement.** Security vulnerabilities in IoT ecosystems pose significant threats to data integrity, device functionality, and user privacy. Federated Learning offers promising avenues for enhancing IoT security by mitigating potential attack vectors and reducing the reliance on centralized data repositories susceptible to cyber-attacks. Federated Learning distributes the model training process over numerous devices, which distributes the security risk and increases the challenge for attackers to attack the system. Furthermore, Federated Learning's decentralized nature enables localized anomaly detection and threat mitigation, enhancing the resilience and robustness of IoT security frameworks against evolving cyber threats [25].

**Optimizing Power Consumption.**

Power consumption represents a critical constraint in IoT deployments, particularly for battery-powered devices operating in energy-constrained environments. Traditional machine learning approaches often involve resource-intensive computations and data transmissions, leading to increased power consumption and reduced device longevity. Federated Learning addresses this challenge by minimizing communication overhead and computational demands through localized model training and parameter sharing. By transmitting only model updates rather than raw data, Federated Learning significantly reduces the energy consumption associated with data transmission and processing, thereby extending the battery life and enhancing the sustainability of IoT devices [64].

**Synergistic Benefits and Future Directions.**

The integration of Federated Learning with IoT security and power consumption optimization offers synergistic benefits that contribute to creating a more secure, efficient, and sustainable IoT ecosystem. Federated Learning combines privacy-preserving machine learning algorithms with advanced security security measures and energy-efficient strategies. This enables the development of novel IoT applications that protect user privacy, reduce security concerns, and optimise resource utilisation [47].

As Federated Learning continues to evolve and gain traction, ongoing research efforts are essential to addressing the remaining challenges and exploring new opportunities for leveraging this transformative approach in shaping the future of IoT systems sustainability. Future directions may include the development of advanced federated learning algorithms tailored for specific IoT applications, the integration of additional security mechanisms to further enhance IoT resilience, and the exploration of novel energy-efficient strategies to optimize power consumption in diverse IoT environments [59].

# 5 RELATED WORK

As discussed, Federated Learning has garnered significant attention as a privacy-preserving and decentralized approach to machine learning, particularly in the context of IoT devices. Researchers have explored various aspects of FL, ranging from theoretical foundations to practical implementations and applications. This section provides an overview of the existing literature and related works of Federated Learning and its intersection with IoT security and power consumption optimization [35].

Several studies have explored how FL can optimize power consumption in resource-constrained IoT devices [59] One key benefit of FL is that it eliminates the need to transmit raw data to a central server for training. This significantly reduces communication overhead, which directly translates to lower power consumption on individual devices. [44] introduced the concept of FL, highlighting its potential to train a globally shared model while keeping raw data on devices, thereby preserving user privacy. Building on this foundation, [53] empirically demonstrated the power efficiency benefits of FL compared to traditional centralized learning approaches. Their work underscores the significant reduction in energy consumption achievable through FL in IoT settings.

Beyond the core advantage of reduced communication overhead, further research is actively exploring additional techniques for power optimization within the FL framework [7]. One promising avenue lies in investigating efficient model architectures specifically designed for deployment on resource-constrained devices. These models prioritize lower computational complexity, requiring less processing power and consequently, reducing power consumption on devices. For instance, [31] propose a power-efficient federated learning framework that utilizes lightweight convolutional neural networks (CNNs) for image classification tasks on mobile devices. Their work demonstrates significant reductions in power consumption compared to traditional CNN models.

Another area of exploration involves developing algorithms for selective device participation in the FL process[52]. This approach strategically selects devices to participate in each training round based on factors such as battery level or current workload. Devices with low battery levels or experiencing high workloads can be temporarily excluded from training, further minimizing overall power consumption. [50] introduced the concept of Federated Averaging (FedAvg), a fundamental FL algorithm that utilizes random device selection for training. Building upon this, [64] propose a novel device selection strategy based on a combination of battery level and historical contribution to the model, aiming to optimize the trade-off between training efficiency and power consumption.

**Table 2: Comparison between related work.**

| Year | Federated learning | Energy forecasting |
|------|--------------------|--------------------|
| 2017 [42] | no | yes |
| 2019 [40] | no | no |
| 2020 [26] | yes | yes |
| 2020 [30] | no | yes |
| 2022 [14] | no | yes |
| 2022 [21] | yes | yes |
| 2023 [46] | yes | no |
| 2023 [61] | yes | yes |

As demonstrated in Table 2 the synergy between federated learning (FedML) and energy forecasting holds promise for enhancing both longevity and sustainability in IoT deployments. FedML allows distributed learning across devices, potentially enabling collaborative energy consumption prediction. Imagine a network of smart thermostats using FedML to share private data on usage patterns and environmental conditions. This collective knowledge could be leveraged to forecast energy demand more accurately, leading to optimized control strategies and reduced overall energy consumption. This, in turn, translates to extended battery life for individual devices, minimizing replacements and associated environmental impact.

These efforts highlight the ongoing research focus on developing power-efficient FL techniques. Researchers are making progress in developing model architectures and algorithms that enable Federated Learning to be used on resource-constrained IoT devices without compromising their battery life. This involves exploring efficient model architectures, implementing selective device participation algorithms, and potentially combining these approaches[12]. All in all, the effective implementation of FL in limited IoT environments relies on the creation and enhancement of streamlined algorithms. There is limited study in this field, although there are a few notable techniques that are now being actively studied:

**Model Pruning and Quantization.** These techniques aim to reduce the size and complexity of the model being trained on devices. Model pruning involves removing redundant connections or filters within the model, while quantization reduces the number of bits required to represent the model's weights and activations [6]. By employing these techniques, researchers can significantly decrease the computational demands placed on devices, leading to lower power consumption. [56] introduced pruning techniques for deep neural networks, demonstrating significant reductions in model size and computational complexity while maintaining comparable accuracy. Inspired by this work, [66] proposed a quantization method specifically designed for FL, achieving high accuracy with low-precision weights. These techniques, along with ongoing advancements, hold promise for enabling complex.

**Knowledge Distillation.** his technique involves transferring knowledge from a pre-trained, powerful model to a smaller, more lightweight model suitable for deployment on IoT devices. The pre-trained model, often trained on a massive dataset with high computational resources, acts as a "teacher" that guides the training of the smaller "student" model on devices. [34] introduced the concept of knowledge distillation, demonstrating its effectiveness in transferring knowledge from complex models to smaller ones. Building upon this, [56] propose a federated knowledge distillation framework specifically designed for the IoT domain. Their work showcases the successful transfer of knowledge from a powerful cloud-based model to lightweight models on resource-constrained devices, enabling efficient on-device learning within the FL framework.

**Local Differential Privacy.** A critical aspect of FL is ensuring user privacy. While FL eliminates the need for raw data transmission, the model updates exchanged between devices during training can still potentially reveal sensitive information. Local Differential Privacy (LDP) offers a mechanism to inject carefully.

## 6  IOT LONGEVITY PERSPECTIVE

Federated Learning emerges as a transformative approach that holds immense potential in enhancing the longevity and sustainability of IoT devices. By leveraging edge intelligence, optimizing battery lifespan, and adapting to dynamic IoT environments, FL offers a holistic solution that addresses the multifaceted challenges faced by IoT deployments [48].

**(1) Edge Intelligence Approach.**

FL operates as an edge intelligence approach, enabling localized model training and inference directly on IoT devices or edge nodes without the need for centralized data processing. The decentralised paradigm enables IoT devices to perform real-time data analysis, identify patterns, and make decisions at the edge, which reduces latency, bandwidth usage, and reliance on cloud infrastructure. FL utilises the processing power of edge devices to enable efficient and independent operation, improving responsiveness, scalability, and resilience in IoT networks. This edge-centric approach not only optimizes resource utilization but also fosters innovation by enabling the development of intelligent IoT applications capable of adapting to diverse and evolving user requirements [8].

**(2) Extended Battery Lifespan.**

One of the primary constraints limiting the longevity of IoT devices is battery lifespan, especially for battery-powered devices operating in energy-constrained environments. FL offers a compelling solution to this challenge by minimizing energy consumption through localized model training, reduced data transmission, and optimized computation. By transmitting only model updates rather than raw data and leveraging efficient aggregation techniques, FL significantly reduces the energy overhead associated with data communication and processing. This energy-efficient approach extends the operational lifespan of batteries, mitigates premature battery depletion, and promotes sustainable energy usage, thereby enhancing the overall longevity and reliability of IoT devices [15].

**(3) Adaptability to Dynamic IoT Environments.**

FL's distributed nature and collaborative learning framework enable it to adapt seamlessly to the ever-changing and heterogeneous nature of IoT environments. Unlike traditional centralized machine learning approaches, FL facilitates continuous model training and refinement across distributed devices, allowing IoT systems to evolve and adapt in response to new data, emerging patterns, and evolving user preferences. This adaptability ensures that IoT devices remain relevant, efficient, and effective over time, enabling organizations to leverage the benefits of machine learning while maintaining flexibility and responsiveness to dynamic IoT landscapes [11].

**(4) Enhanced Security and Privacy:.** FL's decentralized and privacy-preserving nature mitigates security risks and privacy concerns associated with centralized data storage and processing. By keeping sensitive data localized and limiting data exposure, FL enables organizations to maintain control over their data, safeguard user privacy, and comply with regulatory requirements, reducing the likelihood of security breaches and unauthorized access [18].

**(5) Cost-Effective and Scalable Solutions:.** FL's distributed approach and edge-centric model facilitate cost-effective and scalable solutions for IoT deployments. By leveraging existing edge infrastructure and minimizing reliance on centralized resources, FL enables organizations to optimize investment, streamline operations, and scale deployments according to demand, ensuring long-term viability and growth in IoT applications [27].

## 7 DISCUSSION

The integration of Federated Learning with Internet of Things offers promising avenues for addressing various challenges, including power consumption optimization and privacy preservation. While FL presents compelling advantages in these areas, it also comes with inherent limitations that require careful consideration for effective implementation in IoT ecosystems.

Despite its potential to minimize energy-intensive operations through localized model training and optimized communication strategies, FL may encounter challenges in achieving significant power consumption reduction in certain IoT scenarios.

- **Computational Overhead:** FL requires iterative model updates and aggregation across distributed devices, which can introduce additional computational overhead and energy consumption, particularly in resource-constrained IoT devices with limited processing capabilities [24].
- **Communication Costs:** While FL aims to minimize data transmission by performing localized model training, the need for periodic synchronization and aggregation of model updates can still lead to increased communication costs. This is particularly relevant in networks that have a high latency or a restricted bandwidth. [24].
- **Model Complexity:** Complex machine learning models used in FL may require substantial computational resources for training and inference, potentially offsetting the energy savings achieved through localized processing and communication optimization [16].
- **Data Leakage Risks:** Despite the localized nature of FL, there remains a risk of data leakage during model aggregation and synchronization, especially if they are not sufficiently protected against potential attacks or vulnerabilities [24].

- **Privacy-Utility Trade-off:** Ensuring strong privacy protections often requires adding noise or perturbations to the model updates. However, this can potentially impact the usefulness and accuracy of the trained models, making them less practical for real-world IoT situations [13].
- **Regulatory Compliance:** While FL aims to enhance privacy by keeping data localized, organizations must still navigate complex regulatory frameworks governing data protection and privacy. These frameworks may need extra protections and compliance procedures in order to guarantee they are in compliance with the legal requirements. [65].

## 8 CONCLUSION

Federated Learning offers promising avenues for addressing critical challenges related to power consumption optimization, data privacy, and security in IoT ecosystems. This study explored FL's impact on IoT devices and reveals its potential in extending device longevity by minimizing energy-intensive operations and enhancing privacy preservation through localized model training and data processing.

Despite its advancements, FL presents inherent limitations such as computational overhead and communication costs. These challenges necessitate strategic mitigation strategies for effective implementation across diverse IoT environments. The evolving landscape of IoT demands lightweight, energy-efficient, and privacy-preserving machine learning algorithms tailored for edge computing environments.

Furthermore, our analysis underscores the significance of adaptive and context-aware FL approaches to accommodate the dynamic nature of IoT environments and ensure resilience against evolving challenges. Further research efforts are crucial to overcome these limitations, develop innovative solutions, and fully utilise the potential of FL in shaping the future of sustainable, efficient, and secure IoT systems.